\documentclass[aps,prl,twocolumn,groupedaddress]{revtex4}

\usepackage{amsmath}
\usepackage{amsfonts}

\newcommand{\G}{\mathcal{G}}
\newcommand{\R}{\mathbb{R}}

\newcommand{\phat}{\hat{p}}
\newcommand{\qhat}{\hat{q}}
\newcommand{\uu}{\mathbf{u}}
\newcommand{\vv}{\mathbf{v}}
\newcommand{\xx}{\mathbf{x}}
\newcommand{\zz}{\mathbf{z}}

\newcommand{\ket}[1]{\left|#1\right\rangle}

\newcommand{\gen}[1]{\left\langle#1\right\rangle}

\begin{document}

%\preprint{}

\title{Stabilizer Codes for Continuous-variable Quantum Error Correction}


\author{Richard L. Barnes}
\email[]{rbarnes@virginia.edu}
%\thanks{}
\affiliation{Department of Physics, University of Virginia, 382 McCormick Rd., Charlottesville, VA, USA 22904-4714}


\date{\today}

\begin{abstract}
A continuous-variable generalization of the discrete-variable stabilizer formalism for quantum error correcting codes is presented. This generalization is a step toward an independent understanding of continuous-variable quantum information. Our formalism yields all continuous-variable codes discovered to date and can be used to construct continuous-variable analogues of discrete-variable and classical codes.  We use it to rederive the nine-mode code given by Lloyd and Slotine, and a five-mode code obtained by Braunstein. In addition, we construct a new continuous-variable code based on a code of Gottesman mapping three logical modes of information onto eight physical modes.%, and correcting one error. 
\end{abstract}

\pacs{}
%\keywords{}

\maketitle

%\section{I. Introduction }
One of the necessary techniques for the realization of quantum computing is a way to encode and manipulate quantum information in a fault-tolerant manner, so that small, unintended perturbations in the state of system have little or no effect on its large-scale dynamics.  To enable this, quantum error-correcting codes, and in particular, stabilizer codes have been developed.  For systems in $n$-dimensional state space, especially for qubits ($n=2$), general techniques for constructing quantum codes and their associated logical operators are well known.  

Continuous variables are a promising new flavor of quantum information \cite{BP01}, whose potential is still largely unexplored. To date, there have been only a few continuous-variable error-correcting codes constructed.  Lloyd and Slotine have generalized Shor's 9-qubit code \cite{LS98}, and Braunstein constructed a five-mode code, and conjectured about a general technique for converting discrete-variable quantum codes into continuous-variable quantum codes \cite{Bra98a,Bra98b}.  

In this paper, we present such a general technique.  Just as Gottesman \cite{Got96} and Calderbank et al. \cite{CRSS97} showed a correspondence between Pauli operators on $n$ qubits and vectors in $GF(2)^{2n}$, we show an correspondence between operators in the Heisenberg-Weyl (generalized Pauli) group and vectors in $\R^{2n}$.  By rewriting vectors in $GF(2)^{2n}$ as vectors in $\R^{2n}$, we have a general method of constructing continuous variable codes from discrete-variable codes.  As an example, we can reconstruct the examples presented earlier, and present a new code based on a code of Gottesman which encodes three logical qubits in eight physical qubits and corrects one error.

%\section{II. State spaces and Operators }
The Pauli group $G = \{I,X,Y,Z\}$ and the $n$-qubit Pauli groups $G_n = G^{\otimes n}$ play a large role in quantum error correction, because they are a basis for operators on $n$ qubits, and any code that can correct errors in a subset of the Pauli group can then also correct an arbitrary linear combination of such errors \cite{KL97}.  The continuous-variable generalization of the Pauli group is the Heisenberg-Weyl group, or generalized Pauli group.  Let $\qhat$ and $\phat$ be two canonically conjugate observables, $[\qhat,\phat] = 1$, with continuous spectra, for example the position and momentum of a one-dimensional system.  The corresponding state space has as a basis the spectrum of $\qhat$, namely the ``position eigenvectors'' $\ket{q}$, $q \in \R$.  The generalized Pauli group is the collection of all linear shifts in position and momentum.  It is generated by shifts in $p$ and $q$, $X(t) = e^{i\pi t \phat}$ and $Z(t) = e^{i\pi t \qhat}.$  Thus, an arbitrary Pauli operator has the form $X(s) Z(t)$ for some real numbers $s$ and $t$.  The generalized Pauli group acting on $n$ of these systems together is then the collection of tensor products of these operators
\begin{equation} \G_n = \{ X(s_1)Z(t_1) \otimes \dots \otimes X(s_n)Z(t_n) : s_i, t_i \in \R \}. \end{equation}
A typical member of this group can expressed in terms of the canonical observables $q_i$ and $p_i$ as a unitary operator indexed by a vector $\vv = (s_{1},\dots,s_{n},t_{1},\dots,t_{n}) \in \R^{2n}$
\begin{equation} U(\vv) = \exp\left( i \pi \sum_{i=1}^n (s_i \phat_i + t_i \qhat_i) \right), \end{equation}
with any two operators commuting up to a phase:
\begin{equation} U(\vv) U(\vv') = e^{i \pi \omega(\vv,\vv')} U(\vv') U(\vv),\end{equation}
where $\omega(\vv,\vv') = \sum_{i=1}^{n} (s_{i} t'_{i} - s'_{i} t_{i})$ is the standard symplectic form on $\R^{2n}$.  This gives us an immediate correspondence between $\G_n$ and the symplectic vector space $\R^{2n}$: The operator $U(\vv)$ corresponds to the vector $\vv$, multiplication of operators corresponds to addition of vectors (up to phase), and scaling of vectors by real numbers corresponds to taking real-number powers of operators.  

Furthermore, we can use vectors in these symplectic spaces to build up abelian subgroups of the Pauli group: Suppose we have $k$ vectors $\uu_{i},\dots,\uu_{k} \in \R^{2n}$ such that $\omega(\uu_i,\uu_j) = 0$ for all $i,j$.  Then we have a subgroup 
\begin{equation} S = \{ U(\vv) : \vv = \sum_{i=1}^k a_i \uu_i,\ a_i \in \R \} \end{equation} 
of $\G_n$ in one-to-one correspondence with the span of the vectors $\uu_i$. This subgroup $S$ is actually abelian, since any two operators $U(\vv)$ and $U(\vv')$ commute if $\omega(\vv,\vv') = 0$, and if $\vv = \sum_{i=1}^k a_i \uu_i$ and $\vv' = \sum_{i=1}^k a_i' \uu_i$, then 
\begin{equation} \omega(\vv,\vv') = \sum_{ij} a_i a_j' \, \omega(\uu_i,\uu_j) = 0 \end{equation}

%\section{III. Stabilizers Codes for Continuous-Variable Systems }
The construction of stabilizer codes was given by Gottesman in \cite{Got97}: Given a subgroup $S$ of the $n$-qubit Pauli group such that (1) any two elements of $S$ commute, and (2) $-I$ is not in $S$, then the space of states stabilized by $S$ consitutes a nontrivial code 
\begin{equation} C(S) = \{ \ket\psi : M\ket\psi = \ket\psi, \forall M \in S \}. \end{equation}
The same fact is true in continuous dimensions: Any subgroup $S$ of the continuous-variable Pauli group satisfying (1) and (2) stabilizes a nontrivial code.  The case when $S$ is a discrete subgroup was studied by Gottesman et al \cite{GKP01}.  They showed that such a discrete subgroup stabilizes a finite-dimensional subspace of the underlying state space, so that the codes created with such stabilizers cannot be used to encode continuous-variable quantum information.

The stabilizers constructed in \cite{GKP01} correspond to symplectically integral lattices, namely the set of integer linear combinations of $\uu_i \in \R^{2n}$ such that $\omega(\uu_i, \uu_j)$ is an integer.  This criterion is too weak to allow for continuous subgroups, since such subgroups would allow for real linear combinations of the basis vectors.  If, on the other hand, we take any set of $k$ vectors $u_1,\dots,u_k \in \R^{2n}$ satisfying the further restriction that $\omega(\uu_i, \uu_j) = 0$ for all $i,j$, then as described in part II, the subgroup $S$ corresponding to these vectors is an abelian subgroup of the continuous-variable Pauli group $\G_n$, and thus a valid stabilizer.  

Since any operator which commutes with everything in $S$ maps codewords to codewords, the group of such operators, namely the normalizer $N(S)$ of $S$, has a natural interpretation as a collection of operators acting on the code space.  In our symplectic language, an observable commuting with $S$ corresponds to a vector $\vv$ such that $\omega(\vv,\uu_i) = 0$ for all $i$, so that if the $\uu_i$ span a subspace $W \subseteq \R^{2n}$, then the space of operators commuting with $S$ is the symplectic orthogonal $W^\omega = \{ v : \omega(v,\uu_i) = 0 \forall i \}$.  It follows that $\dim W^\omega = \dim \R^{2n} - \dim W = 2n - k, $ so that the group $N(S)/S$ of operators acting nontrivially on the code space corresponds to a subspace of dimension $\dim W^\omega/W = \dim W^\omega - \dim W = 2(n-k)$.  Now we can construct a basis $\xx_1, \zz_1, \ldots, \xx_{n-k}, \zz_{n-k} \in \R^{2n}$ for $W^\omega/W$ such that $\omega(\xx_i,\xx_j) = \omega(\zz_i,\zz_j) = 0$ and $\omega(\xx_i,\zz_j) = \delta_ij$.  The group of operators corresponding to this basis satisfies $[U(s\,\xx_i),U(t\,\xx_j)] = [U(s\,\zz_i),U(t\,\zz_j)] = 0$ and 
\begin{equation}U(s\,\xx_i) U(t\,\zz_j) = e^{i \pi s t \delta_{ij}} U(t\,\zz_j),U(s\,\xx_i),\end{equation} 
and thus has the same group structure as the generalized Pauli group $\G_{n-k}$, with the operators $U(s\,\xx_i)$ and $U(s\,\zz_i)$ acting as position and momentum shifts on the $i$-th logical mode, respectively.

This means that if we can find $k$ vectors $u_1,\dots,u_k \in \R^{2n}$ such that $\omega(\uu_i,\uu_i) = 0$, then we can choose any vector in the code space to represent the $n-k$-mode state $\ket{0\dots0}$, and use the $n-k$ logical operators to transform this logical zero state into an encoding of any other state.  In brief, $k$ symplectically orthogonal vectors give rise to a continuous-variable code encoding $n-k$ continuous-variable systems.

Note that the states which are codewords stabilized by $S$ have certain observable properties: Since they are stable (i.e. eigenvectors with eigenvalue 1) under the action of operators of the form $U(\uu_{j})\exp(\sum_{i} s_{ij} \phat_{i} + t_{ij} \qhat_{i})$, they must be eigenvectors of the observable $\sum_{i} s_{i} \phat_{i} + t_{i} \qhat_{i}$ with eigenvalue zero.  Thus, for each basis vector $\uu_{j} = (s_{1j},\dots,s_{nj},t_{1j},\dots,t_{nj})$ we have an observable $m_{j} = \sum_{i} s_{ij} \phat_{i} + t_{ij} \qhat_{i}$ which is zero on the code space, and conversely, the code space is the set of eigenvectors of these obsevables with eigenvalue zero.  Thus, if we can measure these observables, which are just linear sums of position and momentum, then we can detect whether or not a given state is in the code stabilized by $S$.

%\section{IV. Constructing Codes }
Based on the above, we have a procedure for constructing a continuous-variable quantum code: In order to create a code encoding $n-k$ logical modes into $n$ physical modes of information, we follow the following steps
\begin{enumerate}
\begin{item}Choose $k$ vectors $\uu_1,\dots,\uu_{k} \in \R^{2n}$ such that $\omega(\uu_i,\uu_j) = 0$ for all $i,j$.  
%This defines an abelian subgroup $S$ of $\G_{k}$ with $k$ generators, which stabilizes a nontrivial code space. 
\end{item}
\begin{item}
Compute the standard logical zero state 
\begin{align}
\ket{\bar0} &= (\sum_{M \in S} M) \ket{0\dots0} \\
            &= \int dt_1 \cdots dt_{k} \, U(t_1\, \uu_1) \cdots U(t_{k}\, \uu_{k}) \ket{0\dots0} 
\end{align}
This state is obviously stable under the action of operators in $S$.
\end{item}
\begin{item}
Compute a set of logical Pauli operators.  For these we can take any basis for the symplectic orthogonal $W^\omega$ of subspace $W \subseteq \R^{2n}$ spanned by the vectors $\uu_1,\dots,\uu_{k}$.  
\end{item}
\end{enumerate}

Note that all three of these steps are nearly trivial if we already have a discrete-variable stabilizer code encoding $n-k$ logical qubits into $n$ physical qubits.  In the first step, consider the binary vectors defining the discrete-variable code as real vectors with entries in $\{0,1\} \subseteq \R$, ensuring that $\omega(\uu_i,\uu_j) = 0$, (often possible just by changing some entries from $1$ to $-1$).  The second step is a trivial computation in either case.  In the third step, the binary vectors defining logical operators on the discrete-variable code can again be considered as real vectors, and appropriately modified so that they commute with the continuous-variable stabilizer.  

%\section{V. Some Examples }
%Here we present three examples, each following the method given above for constructing codes.  First we will set up the three-mode position and momentum codes which gives rise to Lloyd and Slotine's generalization of Shor's 9-qubit code \cite{LS98}, then a 5-qubit code presented by Braunstein \cite{Bra98a,Bra98b}, and finally we will present a ew code based on an [[8,3,3]] code presented by Gottesman in \cite{Got96}.

%%%% Three-mode position code
The smallest possible examples of error-correcting codes are the three-mode position and momentum codes mapping $\ket{q}$ to 
\begin{equation} \ket{\bar{q}} = \ket{qqq},\ \ket{\bar{q}} = \int dt_{1}\, dt_{2}\, dt_{3}\, e^{3i \pi (t_{1}+t_{2} + t_{3})q} \ket{t_{1},t_{2},t_{3}}, \end{equation}
and correcting a single shift in $\qhat$ or $\phat$, respectively.  Observe that the position code is invariant under the operators $Z(t)\, Z(-t) \, I$ and $I \, Z(t) \, Z(-t)$:
\begin{equation} (Z(t)\, Z(-t) \, I)\ket{qqq} = e^{i\pi tq} e^{i\pi (-t)q} \ket{qqq} = \ket{qqq}. \end{equation}
Conversely, it is the code stabilized by the group $S= \gen{Z(t)\, Z(-t) \, I, I \, Z(t) \, Z(-t)}$ generated by these operators, which corresponds to the generator matrix
\begin{equation} A = \left( \begin{array}{rrr|rrr}
0 & 0 & 0 & 1 & -1 &  0 \\
0 & 0 & 0 & 0 &  1 & -1 \\
\end{array} \right).\end{equation}  
This is a valid code, since we only have two generators for the stabilizer, corresponding to the rows $\uu_1 = (0,0,0,1,-1,0)$ and $\uu_2 = (0,0,0,0,1,-1)$ of $A$, and $\omega(\uu_1,\uu_2) = 0$.  The logical zero generated in the standard way is also what we expect it to be, namely 
\begin{align}
\ket{\bar0} &= \int dt_1\, dt_2\, e^{i\pi (t_1 - t_2)q} \ket{000} \\
&= \left( \int d(t_1-t_2)\, \delta(t_1-t_2) \right) \ket{000} = \ket{000} 
\end{align}
The standard logical position shift, $\bar{X}(t) = X(t) \otimes X(t) \otimes X(t)$, is still valid here, since its vector $\xx = (1,1,1,0,0,0)$ satisfies $\omega(\xx,\uu_1) = \omega(\xx,\uu_2) = 0$ for the basis vectors $\uu_1$, $\uu_2$ of the stabilizer.  This means that an arbitrary state $\ket{q}$ is encoded as $\bar{X}(q)\ket{\bar0} = (X(t) \otimes X(t) \otimes X(t)) \ket{000} = \ket{qqq}$, which is the standard definition of the three-mode position code.  
%%%% Three-mode position code
A similar construction, reversing the roles of $X$ and $Z$, gives us a three-mode stabilizer code protecting against momentum shifts, and encoding an arbitrary position eigenvector as shown above. 

%%%% Nine-mode Shor code
These two codes can be concatenated in the usual way, by first encoding one qubit into three with the position code, then encoding each of the three bits with the momentum code.  This gives Lloyd and Slotine's generalization of Shor's code, which encodes the state $\ket{q}$ as
\begin{equation} \ket{\bar{q}} = \int dt_i\, e^{i\pi q(t_1+t_2+t_3)} \ket{t_1,t_1,t_1,t_2,t_2,t_2,t_3,t_3,t_3}. \end{equation}
And the logical operators acting on the code, as in the case of the Shor code, are $\bar{X}(t) = (X(t))^{\otimes 9}$ and $\bar{Z}(t) = (Z(t))^{\otimes 9}$.

\begin{table}[b]
$$
A = \left( \begin{array}{rrrrrrrr|rrrrrrrr}
1 & 1 & 1 & 1 & 1 & 1 & 1 & 1 & 0 & 0 & 0 &  0 &  0 &  0 &  0 &  0 \\
0 & 0 & 0 & 0 & 0 & 0 & 0 & 0 & 1 & 1 & 1 &  1 & -1 & -1 & -1 & -1 \\
0 & 1 & 0 & 1 & 1 & 0 & 1 & 0 & 0 & 0 & 0 &  0 & -1 & -1 &  1 &  1 \\
0 & 1 & 0 & 1 & 0 & 1 & 0 & 1 & 0 & 0 & 1 & -1 &  0 &  0 &  1 & -1 \\
0 & 1 & 1 & 0 & 1 & 0 & 0 & 1 & 0 & 1 & 0 & -1 &  0 &  1 &  0 & -1 \\
\end{array} \right).
$$

\begin{eqnarray}
\bar{X}_1(t) &=& X(t) \otimes X(-t) \otimes I     \otimes I     \otimes I    \otimes Z(t)  \otimes I     \otimes Z(-t) \nonumber\\
\bar{X}_2(t) &=& X(t) \otimes I     \otimes X(-t) \otimes Z(-t) \otimes I    \otimes I     \otimes Z(t)  \otimes I     \nonumber\\
\bar{X}_3(t) &=& X(t) \otimes I     \otimes I     \otimes Z     \otimes X(t) \otimes Z(t)  \otimes I     \otimes I     \nonumber\\
\bar{Z}_1(t) &=& I    \otimes Z(-t) \otimes I     \otimes Z(t)  \otimes I    \otimes Z(-t) \otimes I     \otimes Z(t)  \nonumber\\
\bar{Z}_2(t) &=& I    \otimes I     \otimes Z(-t) \otimes Z(-t) \otimes I    \otimes I     \otimes Z(t)  \otimes Z(t)  \nonumber\\
\bar{Z}_3(t) &=& I    \otimes I     \otimes I     \otimes I     \otimes Z(t) \otimes Z(t)  \otimes Z(-t) \otimes Z(-t) 
\end{eqnarray}
\caption{ \label{gotop}
Generator matrix and logical operators for the eight-mode code.
} %% \end{caption}
\end{table}

%%%% Five-mode Braunstein code
Braunstein has presented a five-mode code by showing a network which performs encoding \cite{Bra98a}; this can also be described in this stabilizer formalism.  By examining the logical zero state given by Braunstein, we find that its stabilizer corresponds to the generator matrix
\begin{equation}
A = \left( \begin{array}{rrrrr|rrrrr}
1 & 0 & 0 & -1 & -1 & 0 &  0 & 1 & -1 & 0 \\
0 & 1 & 0 &  0 &  1 & 1 &  0 & 0 &  1 & 0 \\
0 & 0 & 1 &  1 &  0 & 1 &  0 & 0 &  0 & 1 \\
0 & 0 & 0 &  0 &  0 & 0 & -1 & 1 & -1 & 1
\end{array} \right),
\end{equation}
so that we have the encoded zero state
\begin{equation} \ket{\bar0} = \int dt_1\, dt_2\, dt_3\, e^{i \pi t_1 t_2} \ket{t_3, t_2, t_1, t_1-t_3, t_2-t_3}. \end{equation}
For the logical Pauli operators, we take $\bar{X}(t) = Z(t) \otimes X(t) \otimes X(t) \otimes I \otimes I$ (in order to replicate Braunstein's code), and take $\bar{Z}(t) = (F^\dag)^{\otimes 5} \bar{X}(t) (F)^{\otimes 5} = X(t) \otimes Z(t) \otimes Z(t) \otimes I \otimes I$ to be the conjugate of $\bar{X}(t)$ by a fourier transform acting on each mode.  These logical operators correspond to the vectors $\xx = (0,1,1,0,0,1,0,0,0,0)$ and $\zz = (1,0,0,0,0,0,1,1,0,0)$, and thus commute with the stabilizer and form logical Pauli operators since $\omega(\xx,\zz) = 2-1 = 1$.  Acting with $\bar{X}(x)$ on the logical zero state, we find that the encoded version of the state $\ket{x}$ is the logical state
\begin{align}
\ket{\bar{q}} &= (Z(q) \otimes X(q) \otimes X(q) \otimes I \otimes I) \ket{\bar{0}} \\
              &= \int  dt_i\, e^{i \pi (t_1 t_2+t_3q)} \ket{t_3, t_2+q, t_1+q, t_1-t_3, t_2-t_3}, 
\end{align}
which is the same encoded state produced by the network described by Braunstein in \cite{Bra98b}.  Thus, this stabilizer is another way of describing Braunstein's code, with the added advantage of providing the logical operations $\bar{X}(t)$ and $\bar{Z}(t)$.

%%%% Eight-mode stabilizers
%\begin{table}
%\begin{equation}
%\begin{array}{rcccccccccc}
%M_1       &=& X & X & X & X & X & X & X & X \\
%M_2       &=& Z & Z & Z & Z & Z & Z & Z & Z \\
%M_3       &=& I & X & I & X & Y & Z & Y & Z \\
%M_4       &=& I & X & Z & Y & I & X & Z & Y \\
%M_5       &=& I & Y & X & Z & X & Z & I & Y \\ \hline
%\bar{X}_1 &=& X & X & I & I & I & Z & I & Z \\
%\bar{X}_2 &=& X & I & X & Z & I & I & Z & I \\
%\bar{X}_3 &=& X & I & I & Z & X & Z & I & I \\
%\bar{Z}_1 &=& I & Z & I & Z & I & Z & I & Z \\
%\bar{Z}_2 &=& I & I & Z & Z & I & I & Z & Z \\
%\bar{Z}_3 &=& I & I & I & I & Z & Z & Z & Z \\
%\end{array}
%\begin{equation}
%\caption{ \label{gotstab}
%Generators of the stabilizer for Gottesman's 8-qubit code.
%} %% \end{caption}
%\end{table}

%%%% Eight-mode Gottesman code
Finally, to show that this formalism can be used to produce new continuous-variable codes, we can generalize a more interesting code of Gottesman, which encodes 3 logical qubits in 8 physical qubits.  The generator matrix corresponding to this stabilizer, shown in Table \ref{gotop} is identical (except for signs) to the real generator matrix for the continuous-variable code.  The logical Pauli operators follow similarly; by adding sign changes where necessary to the logical operations on the discrete code, we obtain the six operators acting as logical Pauli operators on the encoded codewords, listed in Table \ref{gotop}.  Thus we can write the encoding of any three-mode position eigenstate in terms of these logical operators acting on the logical zero state:
\begin{widetext}
\begin{eqnarray}\ket{\overline{000}} &=& \int dt_{i}\, 
|
	t_1, 
	t_1 +       t_3 + t_4 + t_5 , 
	t_1 +                   t_5 , 
	t_1 +       t_3 + t_4       ,  
	t_1 +       t_3 +       t_5 ,
	t_1 +             t_4       ,
	t_1 +       t_3             ,
	t_1 +             t_4 + t_5  
\rangle \\
\ket{\overline{q_1 q_2 q_3}} &=& (\bar{X}_1(q_1)\, \bar{X}_2(q_2)\, \bar{X}_3(q_3))\ket{\overline{000}} \\
&=&\int dt_{i}\, e^{i\pi(q_{1} t_{5} + q_{2} t_{4} + q_{3} t_{3})}
|
	q_{1} + q_{2} + q_{3} + t_1, 
	t_1 +       t_3 + t_4 + t_5 - q_{1}, 
	t_1 +                   t_5 - q_{2} , \nonumber\\ & & \qquad
	t_1 +       t_3 + t_4       + q_{3} ,  
	t_1 +       t_3 +       t_5 ,
	t_1 +             t_4       ,
	t_1 +       t_3             ,
	t_1 +             t_4 + t_5  
\rangle. 
\end{eqnarray}
\end{widetext}
so that we have a representation of the entire three-mode state space as superpositions of these eight-mode states.  This is the first example of a continuous-variable error-correcting code which encodes more than 1 logical mode of information.

%\section{VI. Conclusions}
Our generalization of Gottesman's stabilizer formalism provides an explicit algorithm for creating continuous-variable quantum codes out of discrete-variable quantum codes.  In particular, it brings the power of the classical theory of error correcting codes to bear on continuous-variable problems, since any code created with the Calderbank-Shor-Steane construction, for example, can be translated into continuous-variable codes as in the above example.

The creation of a large class of error-correcting codes for continuous-variable systems indicates several possibilities.  First is that a high-level symmetry exists between the discrete- and continuous-variable theories of quantum error correction and fault-tolerant computation.  Already in this vein, Bartlett et al.  recently showed in \cite{BSBN02} that a version of the Gottesman-Knill theorem holds for continuous variable systems, which describes how a large class of Hamiltonians in continuous dimension can be effectively simulated on a classical computer, just as in discrete systems.  Second, the combination of the codes as in \cite{GKP01}, encoding discrete systems into continuous systems, with the true continuous codes described above, could make possible much more interesting codes which combine systems of different dimensions (Braunstein conjectured about these in \cite{Bra98b}).  These could, for example, be constructed as the codes stabilized by a subgroup of the continuous Pauli group which is discrete in some dimensions and continuous in others.  

In practice, the approximation of position eigenstates by finitely squeezed states will hinder error-correction using continuous-variable codes.  Braunstein noted in \cite{Bra98b} that this squeezing must make vacuum noise small with respect to both the expected size of errors and relevant length scales the states to be encoded.  Gottesman et al. computed the effect of such dissipative error on their discrete-variable encodings in \cite{GKP01}, but with techniques that do not immediately generalize to continuous codes.  Similar bounds on the fidelity of continuous-variable codes with finite squeezing would allow quantitative measures to be set on the absolute capacity of a continuous-variable channel, and could be the key to a theory of fault-tolerant continuous-variable computation.

\begin{acknowledgments}
We would like to thank Olivier Pfister for his advice and guidance, and Harold Ward and James Rovnyak for their helpful comments.
\end{acknowledgments}

% \bibliography{QIQC}

\end{document}